\documentclass{article}

 \usepackage[preprint]{neurips_2026}

\usepackage[utf8]{inputenc} 
\usepackage[T1]{fontenc}    
\usepackage{hyperref}       
\usepackage{url}            
\usepackage{booktabs}       
\usepackage{amsfonts}       
\usepackage{nicefrac}       
\usepackage{microtype}      
\usepackage{xcolor}         
\usepackage{graphicx}
\usepackage{subcaption}
\usepackage{multirow}
\usepackage{amsmath}

\usepackage{xspace} 

\newcommand{\dset}{LOVW-8.5K\xspace}

\title{Handwriting decoding as a challenging motor task for EEG Foundation Models}

\author{%
  Srinivas Ravishankar$^{1}$ \quad Ishayu Ghosh$^{2}$ \quad Nora Zajzon$^{1}$ \quad Teng Fei$^{1}$ \quad Virginia de Sa$^{1,2}$ \\
  Department of Cognitive Science$^{1}$ \quad Halicioglu Data Science Institute$^{2}$\\
  University of California San Diego\\
  La Jolla, CA, USA \\
  \texttt{srravishankar@ucsd.edu} \\
}

\begin{document}

\maketitle

\begin{abstract}

Recent attempts at creating  Foundation Models (FMs) for Electroencephalography (EEG) have achieved state-of-the-art performance on multiple tasks including Motor Imagery (MI). 
These MI tasks have typically involved coarse classification between imagined limb movements. However, the development of foundation models necessitates diverse datasets, both for pretraining and evaluating the progress of these models.  
In this work, we propose handwriting decoding as a challenging motor task for FMs.  
We show that several existing datasets are potentially confounded, and introduce a dataset that more rigorously evaluates models.
On this dataset, we find that current FMs, despite showing SOTA performance in multiple MI datasets are outperformed by smaller task-specific models.
We also highlight challenges specific to EEG-based handwriting decoding to inform future work.
In our 4-letter classification task, we show that  (a) Knowledge of movement-onset is crucial to reported decoding performance in prior works, with average performance across subjects dropping from $41.3\%$ to $32.4\%$. (b) Increasing test-time signal quality provides significant performance improvements ($45\%$ to $78\%$ in our best subject) compared to scaling training data with single-trial EEG. 
(c) While scaling training data steadily improves decoding performance, existing FMs do not outperform specialist models in handwriting decoding.
We make our code available at \url{https://anonymous.4open.science/r/EEG-Handwriting-BCI-DFCD/}
\end{abstract}

\section{Introduction}
\label{sec:introduction}

Foundation models have achieved remarkable success in vision and language domains. Consequently, there are growing efforts to develop similar models for Electroencephalography (EEG) to advance EEG-based Brain-Computer Interfaces (BCIs). BCIs empower individuals with severe paralysis to communicate by decoding neural activity, significantly enhancing their quality of life \citep{milekovic2018stable, kubler2020history}.

Many existing BCIs utilize motor imagery \citep{decety1996neurophysiological} as a control signal, a paradigm in which the subject attempts or imagines specific motor movements. 
This approach is desirable as it allows for endogenous, asynchronous control, making it an ideal choice for everyday use. 
Classical motor-imagery BCIs have relied on distinct control signals, such as left- versus right-hand clenching, or movements of the tongue, hands, and feet \citep{mohamed2018comparison}. While foundation models have proven effective in tasks requiring such coarse classification, these control signals are inherently constrained by limited Information Transfer Rates (ITR), resulting in slow and laborious communication for the user.



Inspired by the success of handwriting decoding from invasive neural data \cite{willett2021high}, there have been a number of studies attempting to decode handwriting from EEG to improve ITR.
However, the performance of EEG foundation models on this more fine-grained decoding task has not been studied.
We begin by surveying existing datasets for this task.

\subsection{Potential confounds in prior work}

While several EEG studies have attempted to classify handwriting, we find that many of these studies are affected by various confounds owing to their experimental design, leading to systematic overestimation of decoding performance.

\textbf{Oculomotor confounds due to cue design} : One of the earliest studies in EEG-based handwriting decoding \citep{pei2021online} reported an accuracy of 94\% in a 10-way classification between the characters of “HELLO, WORLD!”. However, their experimental design required participants to write the above characters in predefined boxes every trial while looking at the tablet, shown in Appendix Fig. \ref{fig:flawed_exp_design}.
Moreover, eye-movements were not discarded from the EEG. 
In our own analysis, excluding all other predictors, using a single IC representing horizontal eye-movement achieved 85.6\% accuracy in the classification task, while reported accuracy is 94\%. The decoder also failed to generalize across the same letter \texttt{L} in different locations, confirming that it relied heavily on location more than motor patterns (More details on this experiment in Appendix \ref{subsec:oculomotor_confounds}). 
 A later study \cite{kim2024towards} used this dataset and achieved similarly high classification performance $\sim91\%$. It is likely that this work was similarly confounded by eye-movement artifacts owing to the experimental design.

\textbf{Oculomotor confounds due to visual feedback} : A different study \cite{jiang2025neural} achieved $31\%$ across all 26 letters of the English alphabet. In their experimental design, the subject received visual feedback on-screen as they traced the letter.  A natural inclination while writing letters is to follow the pen tip with the eyes, potentially generating oculomotor signals that are characteristic of each letter. The authors find that gamma frequency in the frontal electrodes most contribute to decoding performance, rather than typical motor activity near the midline, suggesting potential oculomotor confounds.

\textbf{Visual and muscle-artifact confounds}:  The highest performance on the entire English alphabet set \citep{tripathi2024neuroair} collected EEG data for all 26 characters of the English alphabet during index-tracing. The reported performance was high, around $45\%$ on the entire letter set. In their experimental design, the onset of the letter cue also acts as the writing cue, as shown in Appendix Fig. \ref{fig:visual_decoding_confound}. Thus, the letter is traced right after it appears on the screen, making any handwriting period overlap with letter-discriminative visual information. Recent studies \cite{lan2023seeingbrainimagereconstruction, fei2024perceptogram} have shown that visual stimuli can be decoded from EEG. 
Thus, decoding performance may also be affected by visual decoding confounds. 
Independently, we find that average decoding performance is $\sim 24\%$, well above chance ($3.8\%$) when using only electrodes \texttt{Fp1}, \texttt{Fp2}, \texttt{T8}, \texttt{TP10}, \texttt{P8} (commonly contaminated with eye/shoulder motion artifacts) as predictors. This indicates that such artifactual information is likely contributing to the high decoding performance. 

\textbf{Temporal confounds due to block design}: Yet other studies are confounded because of block design. EEG is a nonstationary signal, and hence carries strong temporal markers even within a single session. Practically, this means a decoder can distinguish with high accuracy \textit{when} in the session any snippet of EEG was collected, even after preprocessing. Thus, if the classification task is correlated with time (eg. all the samples of a letter are collected together in a block design), a decoder can trivially solve this task, leading to high accuracy. While this has been discussed with respect to visual decoding \cite{li2020perils}, the Mind Pinyin speller \cite{wu2025mind} and 
\cite{sen2025low} suffer from block design in handwriting decoding. 
This is evidenced by the confusion matrix, reproduced in Appendix \ref{subsec:temporal_confounds}. Each letter is most confused with its temporal neighbors, rather than confusions occurring between letters with similar motor patterns as one would expect.

Finally, we identify two works that are potentially free of the above confounds. \cite{crell2024handwritten} reports modest performance, $\sim 26.2\%$ over a 10 letter classification, and their subjects trace the letter inside a box while fixating on the screen, and do not see any visual feedback related to the letter while they write. We also identify \cite{yang2025handwriting}'s experimental design being potentially free of confounds, though the dataset has not been made publicly available yet.
We emphasize that future work must follow the considerations described above to preclude confounds, which arise from experimental design, not from insufficient preprocessing. Critically, preprocessing is not an effective remedy in these cases: once confounded data is recorded, classifiers will almost always exploit residual confound information despite preprocessing.

\vspace{-3mm}
\subsection{Contributions}

A practical system for locked-in patients will require the ability to work with motor imagery, as in the intracortical study, rather than actual motion. While most prior work has dealt with real writing (on a tablet or in the air), our work investigates the unique challenges associated with decoding handwriting in settings where onset timing is unknown, or movement does not occur at all. 


Invasive work \cite{willett2021high} has shown that scaling training data is a crucial step in achieving high performance, and we investigate if an EEG-based decoder similarly scales.
To answer this question, we collect a large number of trials from a subject, to conduct sample-complexity analysis. 
In a complementary analysis, we also investigate if the SNR of a single-trial of EEG acts as a bottleneck to decoding performance.
The contributions of our work are as follows:
\begin{itemize}
    \item Having highlighted significant confounds in existing EEG handwriting studies, we introduce a new dataset \dset free of these confounds, with more than 8.5K trials.
    \item We find that existing EEG FMs are outperformed by specialist models on this dataset, and propose this task as a challenging MI benchmark for upcoming FMs for EEG.
    \item We perform a sample-complexity analysis on this task, revealing that performance improves steadily with increased training data. However, we show that increasing test-time SNR has a much larger effect on increasing performance
    - our best subject increased accuracy from $45\%$ to $78\%$ in 4-way classification.
    \item We find that knowledge of onset timing is crucial to achieve reported decoding performance in the literature; in a realistic setting where onset timing is unknown, performance is drastically reduced, motivating future work aimed at addressing this gap.  
\end{itemize}

\section{Methods}
\label{sec:methods}

\subsection{Datasets}

We use two handwriting datasets in this work. 
The first dataset, \dset, is one we collect ourselves and describe below, which we make public as an evaluation dataset for future FM work. 
The second is the dataset released by \cite{crell2024handwritten}, with a description and supporting experiments provided in the appendix.  
We now describe the \dset dataset in more detail.

Data were collected from four right-handed participants (2 male and 2 female) using 32 EEG channels. All participants provided informed consent and the study received IRB approval (details withheld temporarily for anonymization).  While our cohort size is quite limited compared to prior work \cite{crell2024handwritten}, we opted instead to collect more trials per participant and over multiple days. Our dataset can thus be used to analyze cross-session stability, with some sessions within a participant separated by more than 1 year. A subset of the standard 10-10 montage was used to record more densely from motor areas of the brain, at the expense of fewer electrodes in the occipital and posterior regions of the head, as shown in Figure \ref{fig:custom_montage}. Cz was used as reference during data collection. 



\begin{figure*}
	\centering
	\begin{subfigure}{.25\textwidth}
		\centering
		\includegraphics[width=0.9\linewidth]{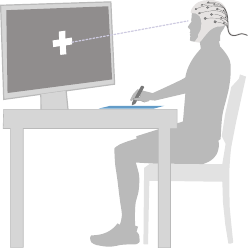}
		\caption{Experimental Setup}
		\label{fig:our_exp_setup}
	\end{subfigure}%
	\begin{subfigure}{.6\textwidth}
		\centering
		\includegraphics[width=0.98\linewidth]{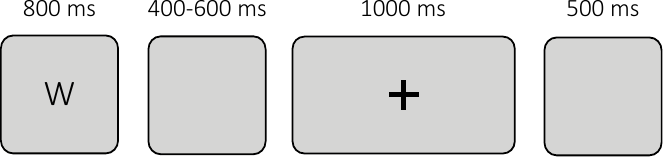}
		\caption{Block diagram}
		\label{fig:block_diagram}
	\end{subfigure}
	\caption{Experimental design for a single trial. The participant fixates on the monitor while writing on the tablet on the desk, to avoid visual feedback. One of the four letters is shown for 800 ms, followed by a blank screen for a randomly chosen period between 400-600 ms. Then a fixation cross appears on the screen for 1000 ms, during which the participant writes the letter on the tablet while looking at the cross. This is followed by a blank screen for 500 ms, after which the next trial begins.}
	\label{fig:experimental_design}
\end{figure*}

To the best of our knowledge, our work is the first to collect handwriting data from participants in both the Motor Execution (ME) and Motor Imagery (MI) paradigms.
In the ME paradigm, participants wrote on an Android tablet placed on the desk in front of them, using a digital stylus as shown in Figure \ref{fig:our_exp_setup}.
Participants were instructed to fixate on the monitor at all times, ensuring that there was no visual feedback that might act as a decoding confound. They practiced writing on the tablet while looking at the computer monitor until they were comfortable. In the Motor Imagery (MI) paradigm, they placed their hands on their lap and imagined writing the letters. 
Four letters were chosen for this study, L, V, O and W, for two reasons: (a) These letters do not require the pen tip to leave the surface of the tablet, thus completely capturing the pen-tip motion corresponding to all muscle movements in the ME paradigm. (b) The letters allow interesting comparative analyses in future work, eg. L and V are rotations, V and W share the first part of their trajectories, etc. There are also a range of discontinuities exhibited by these letters, with W having three sharp discontinuities, L and V having one sharp discontinuity, and O having none.

Each block involved up to 200 trials depending on the participant's fatigue levels. Each trial began with an 800 ms period where the target letter was shown to the participant on the monitor. This was followed by a blank screen for a random period from 400-600 ms. A fixation cross then appeared on screen, which also acted as the go cue for the participant. They wrote or imagined writing the instructed letter within a fixed period of time. In contrast to prior work which provided 4 seconds to write the letter, our movement window is $1000$ ms. We chose such a design to investigate if ballistic imagery can be decoded effectively, potentially improving Information Transfer Rates (ITR). Each trial concluded with a blank period for $500$ ms. Participants were instructed to attempt not to blink during the writing period. The letters were presented in pseudo-random order, with constraints to keep the proportion of different classes balanced, and to prevent consecutive trials from involving the same letter.
The block diagram for a single trial is shown in Figure \ref{fig:block_diagram}.

\begin{figure*}
	\centering
	\begin{subfigure}{0.5\textwidth}
		\centering
		\includegraphics[width=0.9\linewidth]{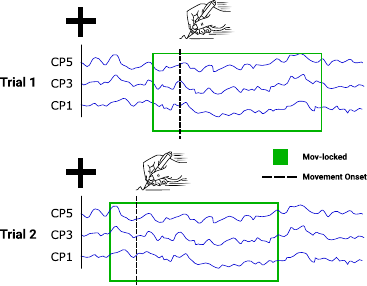}
		\caption{Movement-locked}
		\label{fig:mov_locked}
	\end{subfigure}%
	\begin{subfigure}{0.5\textwidth}
		\centering
		\includegraphics[width=0.9\linewidth]{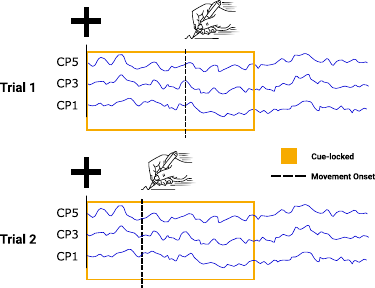}
		\caption{Cue-locked}
		\label{fig:cue_locked}
	\end{subfigure}
	\caption{Epoching settings that represent different difficulties. Movement locked epoching (left) is easier since the movement onset is always a fixed distance from the start of the epoch. However, a realistic setting with imagined handwriting must employ Cue-locked epoching (right) since movement onset is unknown.}
	\label{fig:epoching}
\end{figure*}

Table \ref{tab:data_collection_sched} provides details about the sessions collected over different days, from different participants, and under two possible timing paradigms. The number of trials were equally distributed among the four letters, presented in a pseudo-random order that ensured (a) Each 4 trials had all letters presented (b) No letter was presented twice in succession. 

 We also collected pen-tip trajectories using a custom-built application on the android tablet while the participant was writing the letters. The trajectories were collected at a sampling rate of 200Hz. They were then re-sampled to match the EEG data. Precise synchronization of signals from the EEG amplifier, task markers from the computer and trajectory signals from the tablet was done using photo-diodes, to avoid variations caused by network latency. Further details in Appendix \ref{subsec:synchronization}. 

 The EEG data were collected at 1000 Hz, notch filtered at $60$ Hz, band-pass filtered between $0.3$ Hz and $70$ Hz, decomposed using ICA with ICs representing muscle or eye artifacts  rejected by manual inspection. 
The data were then band-pass filtered between $0.3$ Hz and $40$ Hz, and downsampled to 100 Hz for specialist models, and 200 Hz for foundation models to match pretraining statistics. The data were re-referenced to the Common Average Reference. 
In total, we collected more than 8,500 trials across participants and settings.

\subsection{Decoder}


For our classification decoders, we compare specialist models with foundation models that have been pretrained on large EEG datasets using self-supervision. We evaluate the following commonly used specialist models:
EEGNet \citep{lawhern2018eegnet}, EEGConformer \citep{song2022eeg} and SPaRCNet  \citep{jing2023development},
as well as three recent EEG FMs: CBraMod \citep{wang2024cbramod}, with a Transformer backbone that models spatial and temporal dependencies using parallel attention mechanisms; 
MIRepNet \citep{liu2025mirepnet}, with specialized preprocessing and pretrained specifically for motor tasks, 
and REVE \citep{ouahidi2025reve}, a large-scale FM pretrained on over 60,000 hours of EEG data, with a novel positional encoding scheme enabling it to generalize across electrode configurations.

\subsection{Evaluation}

All models output a letter label corresponding to an epoch of input EEG data of size $C \times T$ where C is the number of EEG channels and T is the temporal length of the epoch.
We quantify the final performance of the system by measuring its accuracy in classifying held-out letter instances as one of four letters (L, V, O and W). The chance performance is $25\%$ since the test data set is balanced. 
We performed nested k-fold cross-validation: 5 folds were created out of each subject's data. For any given test fold, the training fold was independently split again into train and val in a 80/20 ratio, and a limited hyperparameter sweep ($lr : \{1e^{-2}, 1e^{-3}, 1e^{-4}\}$) was conducted. Models were trained for 100 epochs, with the best checkpoint chosen based on validation performance.
To evaluate the model's robustness to non-stationarity, the data were not shuffled. 
This form of evaluation is also essential to prevent the decoder from using time to aid classification (since the temporal blocks between train and test are separate). 
In section \ref{sec:loso_results}, we evaluate in a stricter condition, by leaving each session out entirely during training to measure cross-session decoding stability. 
All models are participant-specific. 

\subsection{Multi-setting Classification}
\label{sec:direct_classification}

We perform classification experiments under 3 settings: ME movement-locked, ME cue-locked, and MI cue-locked. In the first setting, we follow prior work and epoch trials around true movement onset. This is meant as a positive control, since prior work has shown that handwriting in this setting can be successfully decoded. 
In the second setting, we approach a realistic setting where real movement occurs, but precise movement onset is not known to the model. Trials are epoched around the cue to write letters, with timing to actual onset varying across trials.
Finally, the third setting is the most realistic setting possible in healthy participants. Participants were asked to imagine writing the letters while they placed their palms on the lap. Since no actual motion occurred, all trials are cue-locked.

For specialist models, input length was chosen to be 1 second. 
This length was chosen to be longer than the longest duration of letter writing in the data, which was $\sim$ 600 ms. 
For the FMs, input length was chosen through nested cross-validation to be either 4 seconds to match their pretraining statistics, or 1 second to prevent interference from other trials. 

\subsection{Scaling training data vs test-time EEG SNR for performance}



Within the ME movement-locked setting, we further analyze the best performing model to identify what most improves handwriting decoding from EEG: collecting more training data or increasing test-time SNR? Prior efforts in invasive handwriting decoding achieve significant improvements in performance by scaling training data. However, the inherently low SNR of EEG might present a bottleneck. To evaluate this question, we collected large per-class datasets (e.g., $\sim 2400$ trials from participant P1, evenly split across 4 letters).  
To fairly evaluate the effect of training set size, we hold out a fixed test set (the final 160 trials) and train on progressively larger fractions randomly sampled from the remaining data ($10$-$100\%$).  
In parallel, we examine the effect of improving test-time SNR while keeping the training set fixed. Higher effective SNR is simulated by averaging repeated trials of the same class.

\subsection{Neural correlates of handwriting imagery}
\label{sec:explainability}

To ensure that the signal is not confounded, we perform two analyses.
First, we investigate the ERPs corresponding to MI in electrodes at the midline. Concretely, we average all MI cue-locked trials from a session in each participant, and visualize the activity at C1.
We expect the presence of  Movement Related Cortical Potentials (MRCP), which involve  negative deflections corresponding to motor activity  \cite{Shibasaki1980}. 

Second, 
to understand the neural basis of our model's decisions, we employed Gradient-weighted Class Activation Mapping (Grad-CAM) \cite{selvaraju2017grad}, a prominent technique from the explainable AI (XAI) literature.
Grad-CAM is designed to visualize the specific regions of an input trial that influence a classification decision. Briefly, it computes the gradients of a target class score with respect to the feature maps of a final convolutional layer. These gradients are then global-average-pooled to determine the importance weights of each feature map. By taking a weighted combination of these feature maps and applying a ReLU activation, Grad-CAM generates a coarse localization map highlighting the important regions in the input.

We utilize it to analyze a trained EEGNet decoder. With number of channels $C$ and time points $T$, EEGNet treats each input as an image with a single channel of size $C \times T$. We consider one of the trained models from the 5-fold cross validation of participant P1, and average all (N=120) held-out trials of the same class `L' to increase SNR. We confirm that this input is correctly classified as `L', and use Grad-CAM to produce a salience map of the same size $C \times T$ as the input. We average these over time periods of 100 ms, and visualize the salience as a spatiotemporal map on the scalp. 

\vspace{-3mm}
\subsection{Implementation details}
All models were trained on a 4-way classification task using cross-entropy loss with the Adam optimizer. Specialist models were trained from scratch with a learning rate of $1e^{-3}$. FMs were fine-tuned in two configurations: (a) Full Fine-Tuning (Full FT), in which both model backbone and classification head were fine-tuned and (b) Linear Probe (LP), consisting of a frozen backbone with a trainable linear head. In REVE's case, the recommended 2-stage Parameter Efficient FT procedure was followed instead.  All experiments were conducted on a single NVIDIA A6000 GPU.
\section{Results}
\label{sec:results}

\subsection{Foundation models do not outperform specialist models}

Decoding performance of various models on our handwriting dataset is shown in Table \ref{tab:performance_our_dataset}. EEGNet achieved the highest subject-average performance of $41.3\%$. Encouragingly, CBRaMod and REVE show above-chance performance on this out-of-distribution task even with a linear probe. However, they failed to outperform specialist models despite many of the FMs achieving state-of-the-art performance in prior MI datasets. Using a trial-level permutation test described in Appendix \ref{subsec:permutation_test}, subject-average performance differences between EEGNet and the FMs were statistically significant ($p<0.001$) on the \dset dataset.  While our dataset contains only 4 subjects, in a dataset with 20 subjects from prior work \cite{crell2024handwritten}, EEGNet again outperforms REVE FT ($41.1\%$ vs $39\%$ respectively) in a 5-letter handwriting classification task (shown in Appendix Table \ref{tab:crell_reve_vs_eegnet}).  
Overall, these results merit further investigation and suggest handwriting decoding can be used as a challenging motor benchmark for evaluating future FMs. In the next sections, we use the highest performing decoder (EEGNet) to derive insights on the handwriting decoding task.

\begin{table}
\caption{Average decoding performance across subjects in the ME movement-locked setting. (n.s) indicates performance is not significantly above chance accuracy.}
\label{tab:performance_our_dataset}
\vspace{2mm}
\centering
\begin{tabular}{l c c}
\toprule
Model &   Linear Probe & Full FT \\
\midrule
Conformer      & - & 0.345 $\pm$0.037\\
SPaRCNet       & - & 0.359 $\pm$0.052\\
EEGNet         & - & \textbf{0.413 $\pm$0.041}\\
\midrule
MIRepNet       & 0.25 (n.s) & 0.25 (n.s) \\
CBraMod & 0.266 $\pm$0.026 & 0.308 $\pm$0.029 \\
REVE & 0.3057 $\pm$0.037 & 0.322 $\pm$0.049  \\
\bottomrule
\end{tabular}
\end{table}




\subsection{Knowledge of movement onset is crucial to reported performance in prior work}

\begin{figure}
  \begin{subfigure}[h]{0.52\textwidth}
    \centering
    \includegraphics[width=\textwidth]{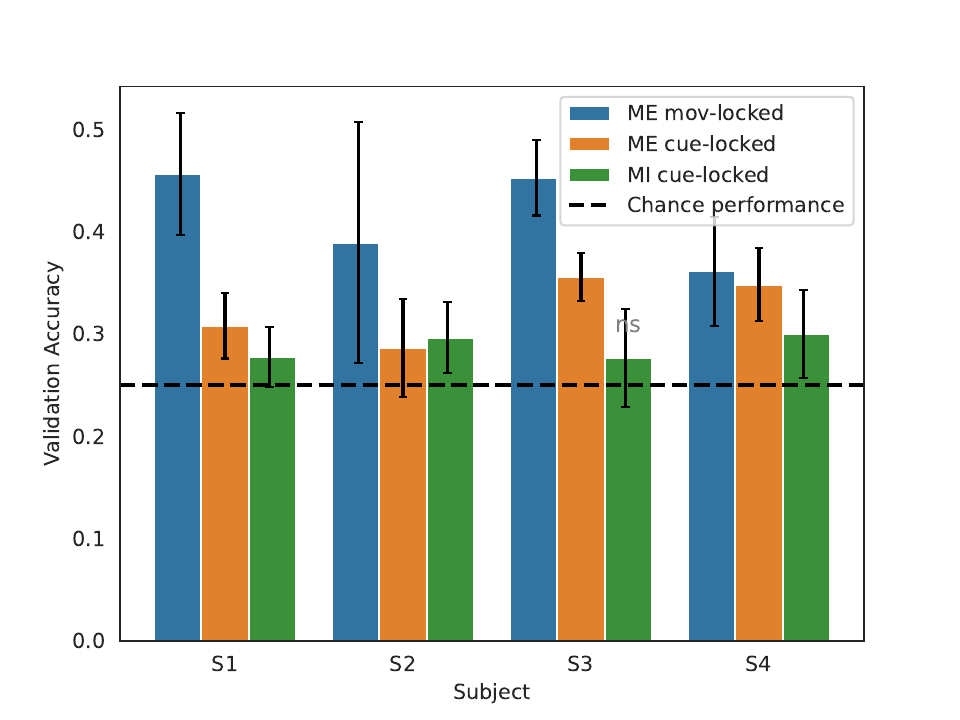}
    \caption{}
    \label{fig:results_consolidated}
  \end{subfigure}
  \hfill
  \begin{subfigure}[h]{0.43\textwidth}
    \centering
    \includegraphics[width=\textwidth]{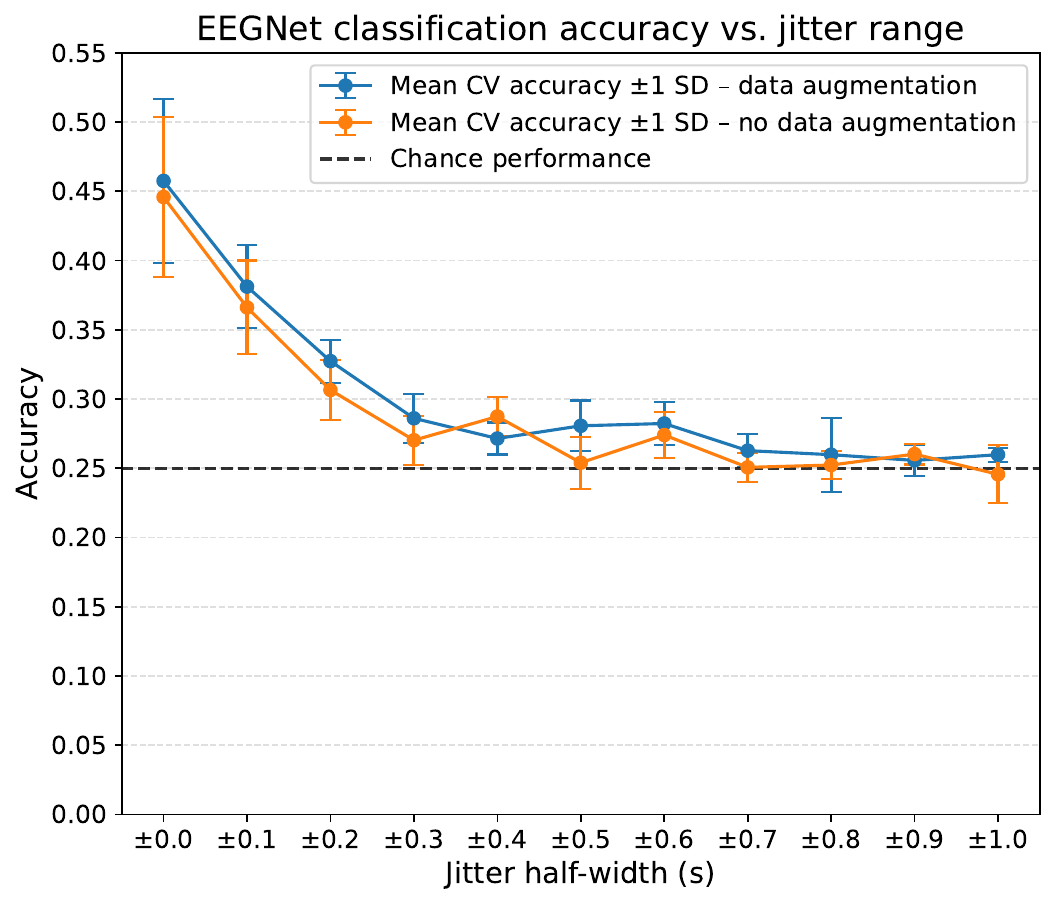}
    \caption{}
    \label{fig:eegnet_jitter_plot}
  \end{subfigure}
  \caption{(a) Participant-wise decoding performance on 3 different settings. Performance drops significantly when motor activity onset is unknown (blue vs orange), indicating a key issue to overcome. Imagined handwriting (green) can be decoded with higher-than-chance performance except for S3. (b) Mean cross-validated classification accuracy (±1 SD across folds) for participant P1 as a function of temporal jitter half-width applied to movement-locked EEG epochs. \textit{Data augmentation} is where each trial is randomly shifted on every access, whereas \textit{no data augmentation} applies a single, trial-specific temporal offset without augmentation.}
\end{figure}

Decoding performance across the three settings and four participants is shown in Figure \ref{fig:results_consolidated}. In line with prior work, all participants achieve higher than chance accuracy in the ME movement-locked setting (blue bar plots), and an overall average of $41.3\% \pm 4.1\%$. 
However, we report a significant drop in performance to $32.45\% \pm 2.86\%$, when the trials are ME cue-locked (orange bar plots) rather than ME movement-locked (blue bar plots).  We also find that this effect is statistically significant ($p<0.001$) at the population level with a Wilcoxon Signed-Rank test (N=$20$) using the handwriting dataset from prior work\cite{crell2024handwritten}, shown in Figure \ref{fig:crell_performance_drop} in the appendix.

To further investigate this drop in decoding performance with variable movement onset, we run a controlled simulation using the ME movement-locked data. The movement-locked epochs are randomly and uniformly jittered around movement onsets with progressively increasing magnitudes, to simulate increasing variability. Figure \ref{fig:eegnet_jitter_plot} shows that performance drops as the onset becomes more variable. Crucially, data augmentation strategies (such as jittering training trials) do not meaningfully mitigate the performance degradation caused by onset uncertainty. Even a small jitter of ±0.2 seconds leads to a drop of over 10\% in accuracy. Although random temporal shifts during training produce a marginally higher accuracy than fixed misalignment, the two curves exhibit nearly identical drop-off behavior as jitter increases.
Complementary analysis can be found in Appendix \ref{subsec:reaction_time_variability}.

Finally, we show that imagined handwriting (MI) can be decoded with greater than chance performance, at $28.758 \pm 1.064$ (in all participants except P3). While performance in this setting is also lower than the movement-locked ME setting, part of this drop is likely explained by what the observed drop in performance when onset information is unknown. FMs have traditionally achieved SOTA performance on coarse-grained MI tasks even with unknown onset information, further motivating the inclusion of fine-grained handwriting decoding in future FM benchmarks.


\subsection{Scaling training data improves performance, but single-trial EEG SNR is a performance bottleneck}
\label{subsec:single_trial_SNR}



\begin{figure}
    \centering
    \includegraphics[width=0.6\textwidth]{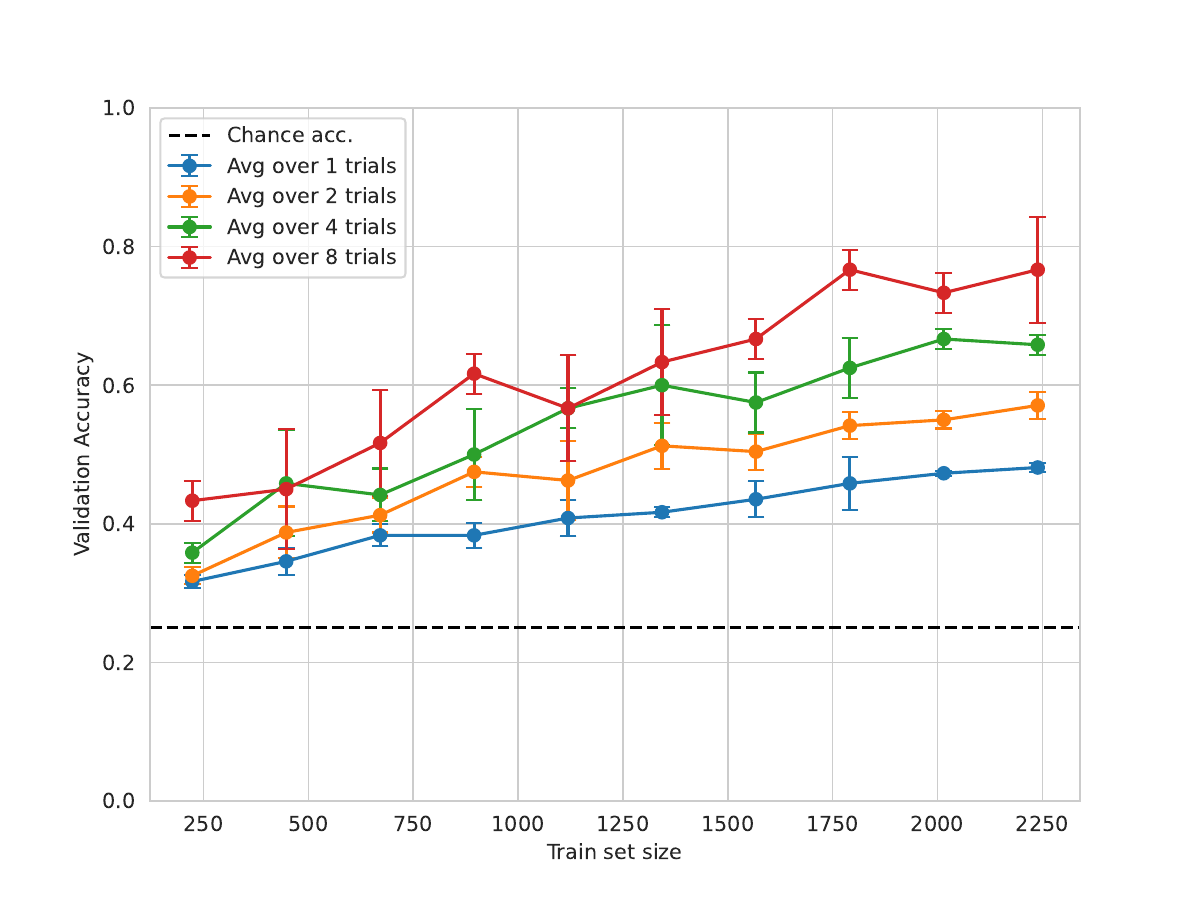}
	\caption{Decoding performance on P1's last 160 ME movement-locked trials, when scaling training data and averaging over various number of trials of the same letter. Mean and standard deviation over 3 random seeds. While there is some performance improvement when scaling training data, boosting test-time SNR significantly improves performance even when training with single-trial EEG }
    \label{fig:SNR_boosted}
\end{figure}

We evaluated the effects of (a) Varying training dataset size and (b) test-time SNR on decoding performance. Test-time SNR was increased by averaging multiple trials of the same letter (aligned to movement onset) during evaluation.
As shown in Figure \ref{fig:SNR_boosted}, scaling the training set yields modest single-trial accuracy gains, but boosting test-time SNR produces far larger improvements. In P1, accuracy rose from 45\% on single trials to 78\% when averaging eight trials.

\subsection{Neural correlates of handwriting}

\begin{figure}
  \begin{subfigure}{0.4\textwidth}
    \centering
    \includegraphics[width=\textwidth]{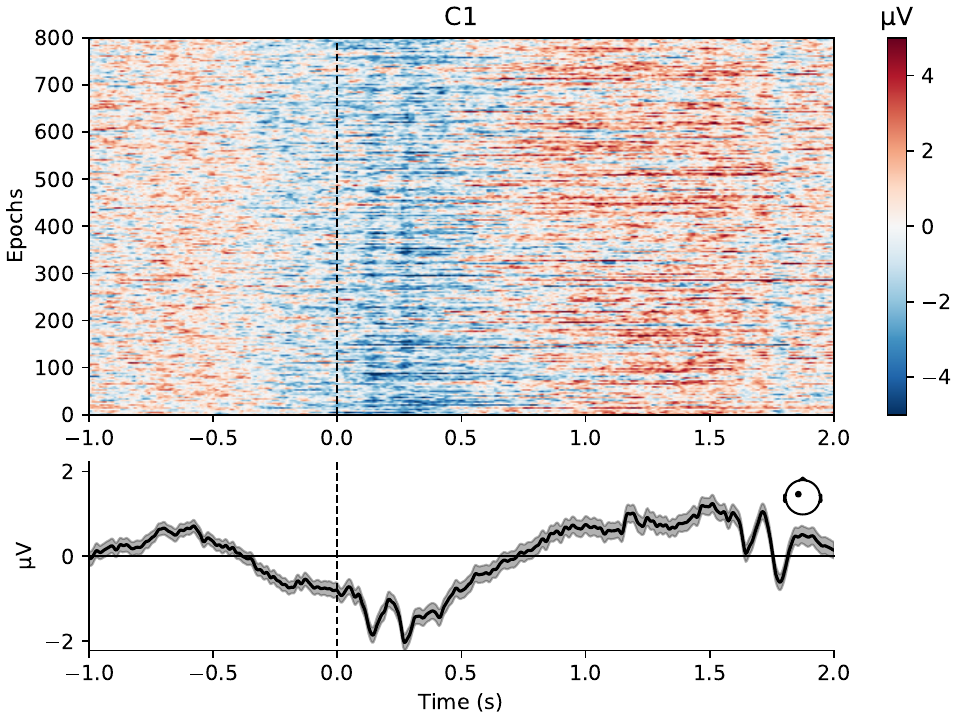}
  \end{subfigure}
  \hfill
  \begin{subfigure}{0.54\textwidth}
    \centering
    \includegraphics[width=\textwidth]{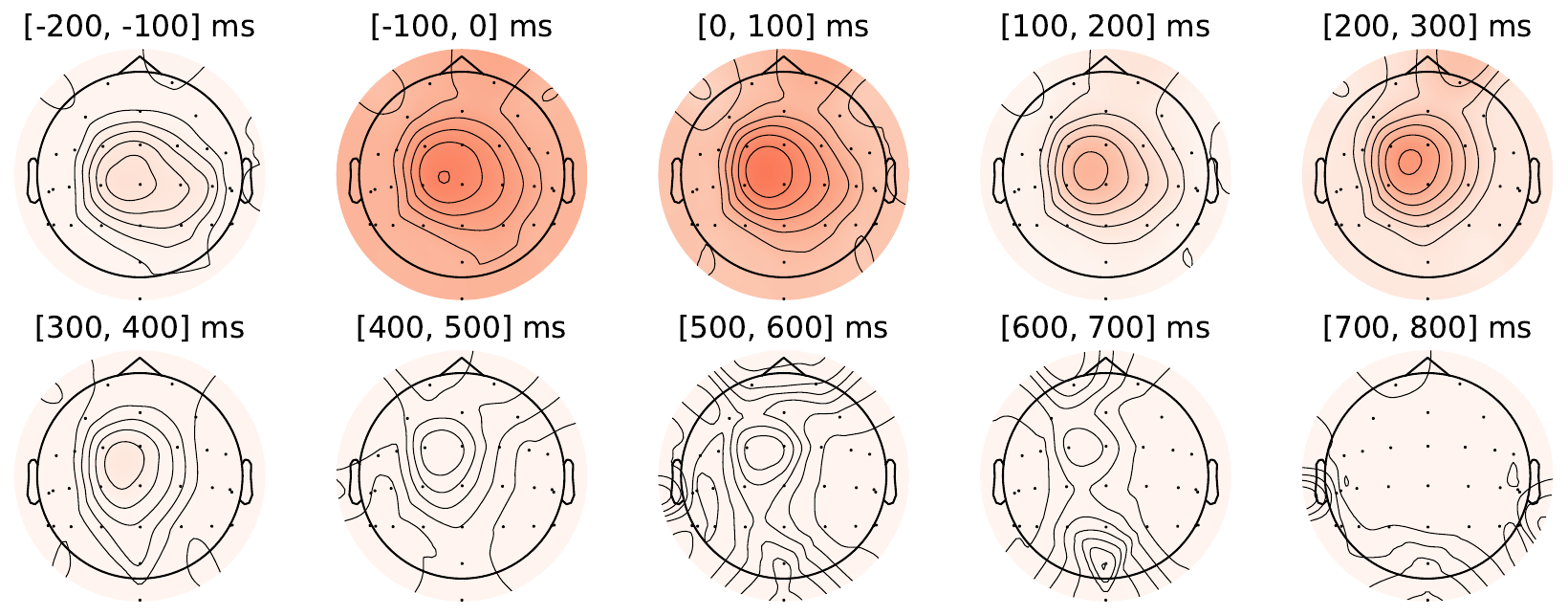}
  \end{subfigure}
  \caption{(a): Movement Related Cortical Potential (MRCP) signal during Handwriting Imagery (MI) at electrode C1. Dashed line represents go cue, upon which participant imagined writing letters. (b) : GradCAM-derived topographic scalp maps indicating input salience for Motor Imagery (MI) trials of letter L in participant P1.} 
  \label{fig:interpretability}
\end{figure}

In this section, we confirm the presence of expected neural signals within the data, and investigate the signal components that contribute most to decoding performance.
We first compute the averaged ERP for all MI trials from the first session of participant P1, at electrode C1. We expect a negative deflection at this site, which is contralateral to the right hand that was used to perform the motor imagery. The stacked trials and averaged waveform is shown in Figure \ref{fig:interpretability}, which aligns with prior work on the neural correlates of motor activity in EEG \cite{Shibasaki1980}.

Next, we investigate a trained EEGNet decoder in the MI setting using the GradCAM procedure described in section \ref{sec:explainability}, identify which parts of the input were used to make classification decisions. The spatiotemporal salience map from [-200 ms, 800 ms] around the fixation/go cue is visualized as a topographic scalp map in Figure \ref{fig:interpretability}. We make two observations: Spatially, Grad-CAM assigns salience to the midline area roughly in the left hemisphere, confirming that the decoder uses the signal described in the previous analysis. More interestingly, salience is assigned to the brain data \textit{pre-fixation} from [-100, 0] ms in addition to the periods after the go-cue, indicating that Motor Imagery planning or intention contribute to decoding performance.  

\subsection{Cross-session decoding performance above chance}
\label{sec:loso_results}

In this section, we examine decoding stability (leave-one-session-out evaluation) across 4 sessions in participant P1. The decoding performance in the movement-locked ME setting is shown in Table \ref{tab:loso_results}. In all sessions, though some transfer learning occurs, performance dropped from the $45\%$ achieved in P1 when within-session examples are included in training. However, the MI setting showed no above-chance cross-session decoding performance, shown in Appendix Table \ref{tab:loso_MI_results}.

\begin{table}
  \caption{Leave-one-session-out accuracies for ME movement-locked for P1 and P2.
           Values are reported as accuracy $\pm$ standard error (SE). Performance is consistently above chance.}
  \label{tab:loso_results}
  \begin{center}
    \begin{small}
      \begin{sc}
        \begin{tabular}{lrrl rr}
          \toprule
          & \multicolumn{2}{c}{P1} & & \multicolumn{2}{c}{P2} \\
          \cmidrule(lr){2-3} \cmidrule(lr){5-6}
          Session & \# Trials & Accuracy & & \# Trials & Accuracy \\
          \midrule
          Session 1 & 315  & $0.327\pm0.026$ & & 339 & $0.286\pm0.025$ \\
          Session 2 & 696  & $0.378\pm0.018$ & & 336 & $0.307\pm0.025$ \\
          Session 3 & 1268 & $0.364\pm0.014$ & & 778 & $0.301\pm0.016$ \\
          Session 4 & 120  & $0.425\pm0.045$ & & 319 & $0.273\pm0.025$ \\
          \bottomrule
        \end{tabular}
      \end{sc}
    \end{small}
  \end{center}
\end{table}

\section{Discussion}
\label{sec:discussion}

To summarize, handwriting decoding represents a potentially challenging task that can drive progress in FMs for EEG. We showed that multiple prior works in handwriting decoding suffer from potential confounds, and make recommendations in experimental design to avoid them in future work. 
We collected a dataset that attempts to avoid the prior confounds, and study decoding performance across a number of FMs and specialist models.

Our results support recent work showing that motor information associated with handwriting can be decoded from EEG. Average single-trial decoding performance over subjects was $41.3\%$ in a 4-way letter classification. Importantly, we show that the precise timing of movement onset is crucial to achieving this performance. With real motion but unknown movement onset, performance dropped significantly. Crucially, this was not improved with data augmentation strategies that attempted to make the model temporally invariant. This may suggest that the motion-onset EEG activity is not clearly detectable in the noisy signal. 

Additionally, this work demonstrated that decoding handwriting from purely imagined handwriting is possible, but exhibits lower performance compared to movement-centered decoding with real movement. Notably, the decoding performance is on par with the cue-centered ME trials, suggesting that solutions to the onset detection problem might result in  performance gains.
Our sample complexity results show a promising trend of improved performance with increased training data, indicating that this task could benefit greatly from FM training.
However, using our dataset, we showed that current FMs do not outperform non-pretrained models despite achieving SOTA performance in other MI tasks. 
This suggests that FMs for EEG might benefit from more diverse pretraining. 

We hypothesize that MIRepNet's chance performance is due to the strict filtering (8-30 Hz) in their pretraining. Prior work \cite{crell2024handwritten} observed useful signals for decoding handwriting in the 0.3-3Hz range, suggesting that FMs should employ minimal preprocessing during pretraining in anticipation of tasks that are outside the domain of the training data.

Test-time SNR remains a key bottleneck, with significant improvements obtained from averaging time-locked evaluation trials. However, it remains to be seen if such a solution leads to similar gains in a practical online setting, where movement onset is unknown. Solutions that involve repetition of stimuli must account for the suppression in neural activity that occurs due to habituation.
Our results motivate training FMs on data spanning multiple SNR levels to increase their denoising capability.


Data from one of our participants involved sessions spanning over a year, allowing us to analyze long-term stability. Leave-one-session-out evaluation across 4 sessions showed successful transfer. 

We make the our dataset publicly available to expand the scope of evaluation for EEG Foundation Models, and inform future work on the handwriting decoding task. 



\bibliography{references}

@article{Shibasaki1980,
  title = {Components of the movement-related cortical potential and their scalp topography},
  volume = {49},
  ISSN = {0013-4694},
  url = {http://dx.doi.org/10.1016/0013-4694(80)90216-3},
  DOI = {10.1016/0013-4694(80)90216-3},
  number = {3–4},
  journal = {Electroencephalography and Clinical Neurophysiology},
  publisher = {Elsevier BV},
  author = {Shibasaki,  H and Barrett,  G and Halliday,  Elise and Halliday,  A.M},
  year = {1980},
  month = aug,
  pages = {213–226}
}

@inproceedings{selvaraju2017grad,
  title={Grad-cam: Visual explanations from deep networks via gradient-based localization},
  author={Selvaraju, Ramprasaath R and Cogswell, Michael and Das, Abhishek and Vedantam, Ramakrishna and Parikh, Devi and Batra, Dhruv},
  booktitle={Proceedings of the IEEE international conference on computer vision},
  pages={618--626},
  year={2017}
}

@article{li2020perils,
  title={The perils and pitfalls of block design for {EEG} classification experiments},
  author={Li, Ren and Johansen, Jared S and Ahmed, Hamad and Ilyevsky, Thomas V and Wilbur, Ronnie B and Bharadwaj, Hari M and Siskind, Jeffrey Mark},
  journal={IEEE Transactions on Pattern Analysis and Machine Intelligence},
  volume={43},
  number={1},
  pages={316--333},
  year={2020},
  publisher={IEEE}
}

@article{yang2025handwriting,
  title={Handwriting Imagery {EEG} Classification based on Convolutional Neural Networks},
  author={Yang, Hao and Ouyang, Guang},
  journal={arXiv preprint arXiv:2509.03111},
  year={2025}
}

@article{ouahidi2025reve,
  title={{REVE}: A Foundation Model for {EEG}--Adapting to Any Setup with Large-Scale Pretraining on 25,000 Subjects},
  author={Ouahidi, Yassine El and Lys, Jonathan and Th{\"o}lke, Philipp and Farrugia, Nicolas and Pasdeloup, Bastien and Gripon, Vincent and Jerbi, Karim and Lioi, Giulia},
  journal={arXiv preprint arXiv:2510.21585},
  year={2025}
}

@article{song2022eeg,
  title={{EEG} conformer: Convolutional transformer for {EEG} decoding and visualization},
  author={Song, Yonghao and Zheng, Qingqing and Liu, Bingchuan and Gao, Xiaorong},
  journal={IEEE Transactions on Neural Systems and Rehabilitation Engineering},
  volume={31},
  pages={710--719},
  year={2022},
  publisher={IEEE}
}

@article{wu2025mind,
  title={Mind-Pinyin Speller: A Non-Invasive Brain-Computer Interface for Efficient Chinese Character Input Using {EEG}-Based Imagined Handwriting},
  author={Wu, Lingyu and Jung, Tzyy-Ping and Li, Xiaojian and Zhou, Yanhong and Wan, Xianglong and Jiao, Wenlong and Xie, Xueguang and Duan, Dingna and Liu, Tiange and Yu, Hao and others},
  journal={Expert Systems with Applications},
  pages={129999},
  year={2025},
  publisher={Elsevier}
}

@article{sen2025low,
  title={A low-latency neural inference framework for real-time handwriting recognition from {EEG} signals on an edge device: O. Sen et al.},
  author={Sen, Ovishake and Soni, Raghav and Virmani, Darpan and Parekh, Akshar and Lehman, Patrick and Jena, Sarthak and Katikhaneni, Adithi and Khalifa, Adam and Chatterjee, Baibhab},
  journal={Scientific Reports},
  volume={15},
  number={1},
  pages={41040},
  year={2025},
  publisher={Nature Publishing Group UK London}
}

@article{wang2024cbramod,
  title={{CB}ra{M}od: A criss-cross brain foundation model for {EEG} decoding},
  author={Wang, Jiquan and Zhao, Sha and Luo, Zhiling and Zhou, Yangxuan and Jiang, Haiteng and Li, Shijian and Li, Tao and Pan, Gang},
  journal={arXiv preprint arXiv:2412.07236},
  year={2024}
}

@article{liu2025mirepnet,
  title={{MIRepNet}: A Pipeline and Foundation Model for {EEG}-Based Motor Imagery Classification},
  author={Liu, Dingkun and Chen, Zhu and Luo, Jingwei and Lian, Shijie and Wu, Dongrui},
  journal={arXiv preprint arXiv:2507.20254},
  year={2025}
}

@article{fei2024perceptogram,
  title={Perceptogram: Reconstructing Visual Percepts from {EEG}},
  author={Fei, Teng and Uppal, Abhinav and Jackson, Ian and Ravishankar, Srinivas and Wang, David and de Sa, Virginia R},
  journal={arXiv preprint arXiv:2404.01250},
  year={2024}
}

@misc{lan2023seeingbrainimagereconstruction,
      title={Seeing through the Brain: Image Reconstruction of Visual Perception from Human Brain Signals}, 
      author={Yu-Ting Lan and Kan Ren and Yansen Wang and Wei-Long Zheng and Dongsheng Li and Bao-Liang Lu and Lili Qiu},
      year={2023},
      eprint={2308.02510},
      archivePrefix={arXiv},
      primaryClass={eess.IV},
      url={https://arxiv.org/abs/2308.02510}, 
}

@article{kim2024towards,
  title={Towards Scalable Handwriting Communication via {EEG} Decoding and Latent Embedding Integration},
  author={Kim, Jun-Young and Kim, Deok-Seon and Lee, Seo-Hyun},
  journal={arXiv preprint arXiv:2411.09170},
  year={2024}
}

@article{jiang2025neural,
  title={Neural Spelling: A Spell-Based {BCI} System for Language Neural Decoding},
  author={Jiang, Xiaowei and Zhou, Charles and Duan, Yiqun and Zhao, Ziyi and Do, Thomas and Lin, Chin-Teng},
  journal={arXiv preprint arXiv:2501.17489},
  year={2025}
}

@article{willett2021high,
  title={High-performance brain-to-text communication via handwriting},
  author={Willett, Francis R and Avansino, Donald T and Hochberg, Leigh R and Henderson, Jaimie M and Shenoy, Krishna V},
  journal={Nature},
  volume={593},
  number={7858},
  pages={249--254},
  year={2021},
  publisher={Nature Publishing Group UK London}
}

@article{pei2021online,
  title={Online recognition of handwritten characters from scalp-recorded brain activities during handwriting},
  author={Pei, Leisi and Ouyang, Guang},
  journal={Journal of Neural Engineering},
  volume={18},
  number={4},
  pages={046070},
  year={2021},
  publisher={IOP Publishing}
}

@article{tripathi2024neuroair,
  title={{NeuroAiR}: Deep Learning Framework for Airwriting Recognition from Scalp-recorded Neural Signals},
  author={Tripathi, Ayush and Gupta, Aryan and Prathosh, AP and Muthukrishnan, Suriya Prakash and Kumar, Lalan},
  journal={IEEE Transactions on Instrumentation and Measurement},
  year={2024},
  publisher={IEEE}
}

@article{lawhern2018eegnet,
  title={{EEGNet}: a compact convolutional neural network for {EEG}-based brain-computer interfaces},
  author={Lawhern, Vernon J and Solon, Amelia J and Waytowich, Nicholas R and Gordon, Stephen M and Hung, Chou P and Lance, Brent J},
  journal={Journal of neural engineering},
  volume={15},
  number={5},
  pages={056013},
  year={2018},
  publisher={iOP Publishing}
}

@article{milekovic2018stable,
  title={Stable long-term {BCI}-enabled communication in {ALS} and locked-in syndrome using {LFP} signals},
  author={Milekovic, Tomislav and Sarma, Anish A and Bacher, Daniel and Simeral, John D and Saab, Jad and Pandarinath, Chethan and Sorice, Brittany L and Blabe, Christine and Oakley, Erin M and Tringale, Kathryn R and others},
  journal={Journal of neurophysiology},
  volume={120},
  number={7},
  pages={343--360},
  year={2018},
  publisher={American Physiological Society Bethesda, MD}
}

@article{kubler2020history,
  title={The history of {BCI}: From a vision for the future to real support for personhood in people with locked-in syndrome},
  author={K{\"u}bler, Andrea},
  journal={Neuroethics},
  volume={13},
  number={2},
  pages={163--180},
  year={2020},
  publisher={Springer}
}

@article{decety1996neurophysiological,
  title={The neurophysiological basis of motor imagery},
  author={Decety, Jean},
  journal={Behavioural brain research},
  volume={77},
  number={1-2},
  pages={45--52},
  year={1996},
  publisher={Elsevier}
}

@article{mohamed2018comparison,
  title={Comparison of {EEG} signal decomposition methods in classification of motor-imagery {BCI}},
  author={Mohamed, Eltaf Abdalsalam and Yusoff, Mohd Zuki and Malik, Aamir Saeed and Bahloul, Mohammad Rida and Adam, Dalia Mahmoud and Adam, Ibrahim Khalil},
  journal={Multimedia Tools and Applications},
  volume={77},
  pages={21305--21327},
  year={2018},
  publisher={Springer}
}

@article{crell2024handwritten,
  title={Handwritten character classification from {EEG} through continuous kinematic decoding},
  author={Crell, Markus R and M{\"u}ller-Putz, Gernot R},
  journal={Computers in Biology and Medicine},
  volume={182},
  pages={109132},
  year={2024},
  publisher={Elsevier}
}

@article {HBM:HBM23730,
author = {Schirrmeister, Robin Tibor and Springenberg, Jost Tobias and Fiederer,
  Lukas Dominique Josef and Glasstetter, Martin and Eggensperger, Katharina and Tangermann, Michael and
  Hutter, Frank and Burgard, Wolfram and Ball, Tonio},
title = {Deep learning with convolutional neural networks for EEG decoding and visualization},
journal = {Human Brain Mapping},
issn = {1097-0193},
url = {http://dx.doi.org/10.1002/hbm.23730},
doi = {10.1002/hbm.23730},
month = {aug},
year = {2017},
}

@article{jing2023development,
  title={Development of expert-level classification of seizures and rhythmic and periodic patterns during {EEG} interpretation},
  author={Jing, Jin and Ge, Wendong and Hong, Shenda and Fernandes, Marta Bento and Lin, Zhen and Yang, Chaoqi and An, Sungtae and Struck, Aaron F and Herlopian, Aline and Karakis, Ioannis and others},
  journal={Neurology},
  volume={100},
  number={17},
  pages={e1750--e1762},
  year={2023},
  publisher={Lippincott Williams \& Wilkins Hagerstown, MD}
}
\bibliographystyle{unsrt}


\newpage
\appendix

\section{Confounded Experimental Designs}
\label{sec:confound_supplementary}

\subsection{Oculomotor confounds}
\label{subsec:oculomotor_confounds}

\begin{figure}[h]
	\centering
	\begin{subfigure}{.45\textwidth}
		\centering
		\includegraphics[width=0.9\linewidth]{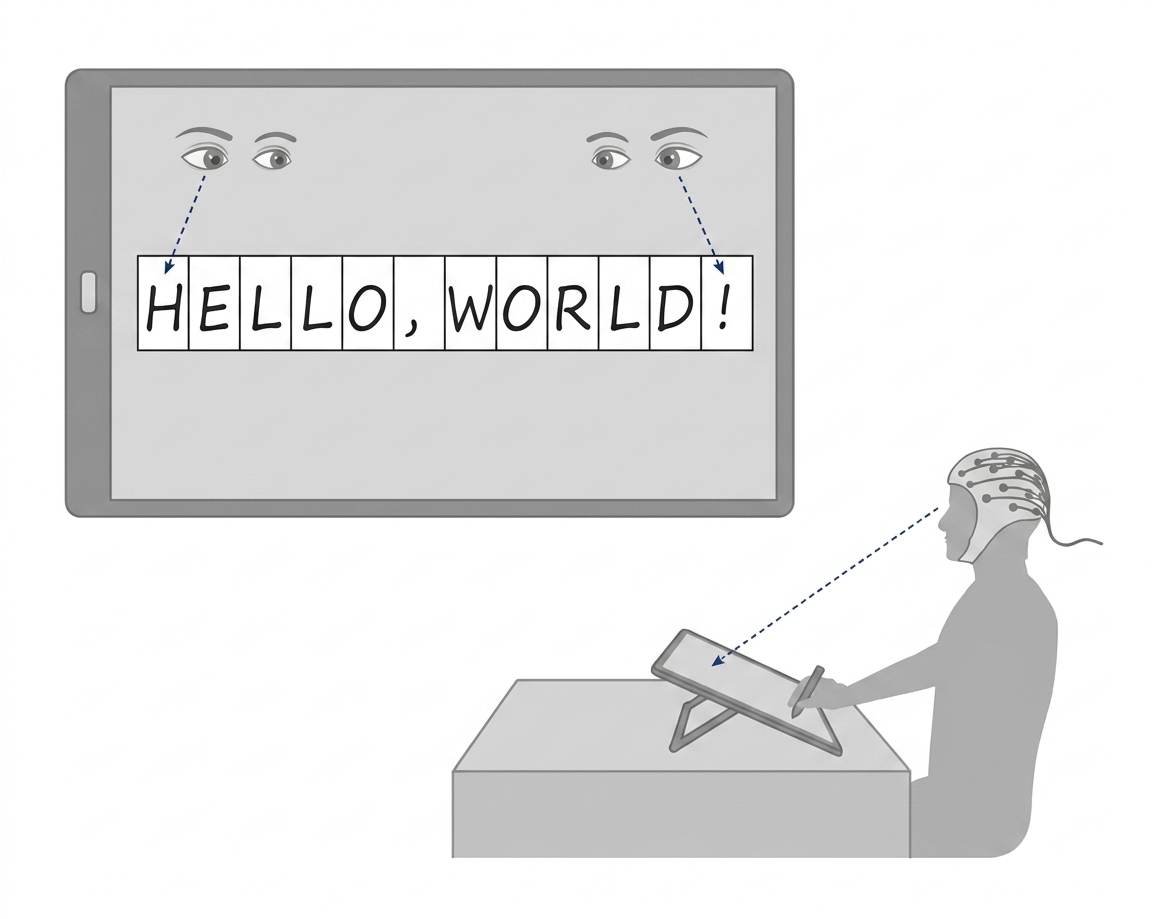}
		\caption{Prior work's exp. design}
		\label{fig:flawed_exp_design}
	\end{subfigure}%
	\begin{subfigure}{.5\textwidth}
		\centering
		\includegraphics[width=0.7\linewidth]{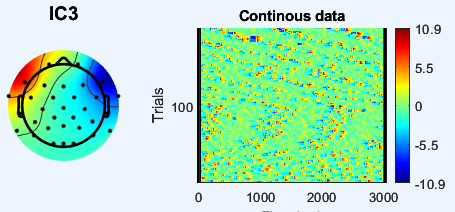}
		\caption{Oculomotor IC}
		\label{fig:eye_IC}
	\end{subfigure}
	\caption{(a) Flawed exp. design, with horizontal eye motion predictive of the letter being written. (b) One of the 10 ICs used for prediction in prior work. The IC represents the horizontal motion of the eyes tracking the letters in different boxes. This IC alone achieves 85.6\% accuracy.}
	\label{fig:prior_work}
\end{figure}


With the oculomotor IC in Fig \ref{fig:eye_IC} as the sole predictor, we were able to achieve a high (85.6\%) `handwriting' decoding performance. Oculomotor artifacts were not removed in the original study.
We also conducted an experiment where the decoder only saw EEG snippets corresponding to letter \texttt{L} from the first word during training, along with all other letters. During evaluation, all instances of \texttt{L} came from the second word. This experiment is illustrated further in \ref{eq:spatial_generalization}. The decoder failed to achieve above-chance performance in classifying EEG snippets from a different position, despite an identical motor pattern seen during training.

\begin{align}
\label{eq:spatial_generalization}
    \text{Training} : \text{H E \textcolor{blue}{L} \textcolor{blue}{L} O , W O R L D !} \\
    \text{Validation} : \text{H E L L O , W O R \textcolor{blue}{L} D !}
\end{align}




\subsection{Visual decoding confounds}

\begin{figure}[h!]
    \centering
    \includegraphics[width=0.7\textwidth]{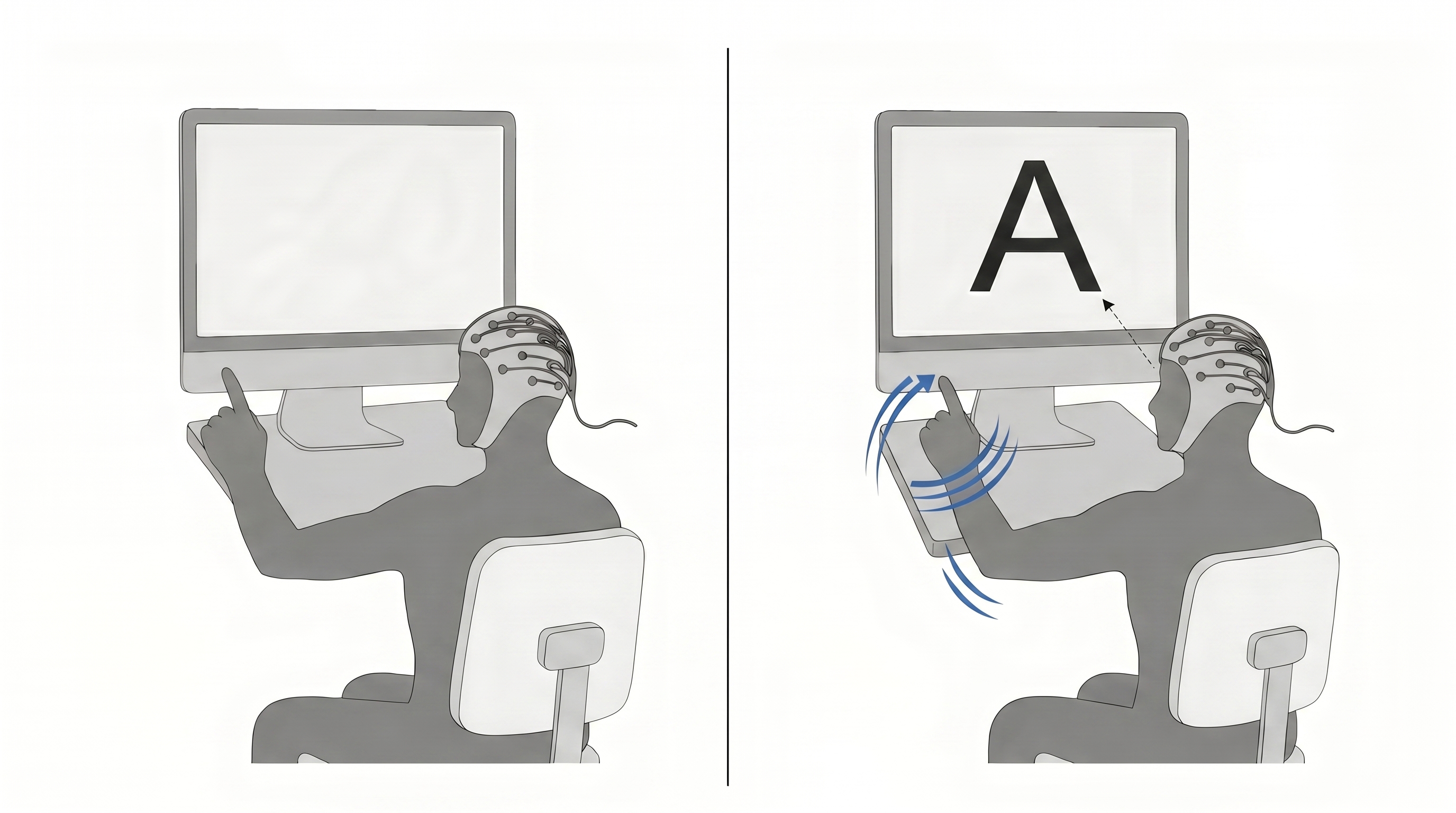}
	\caption{Experimental design with potential visual decoding confounds. Participant is instructed to begin writing the letter as soon as it's cued. This makes the handwriting decoding period involve letter-discriminative visual perception information. Recent work~\cite{fei2024perceptogram} has shown that complex visual stimuli can be decoded from EEG.}
    \label{fig:visual_decoding_confound}
\end{figure}

\subsection{Temporal confounds from block design}
\label{subsec:temporal_confounds}

Under a block design all trials from the same letter occur within the same period of the recording session. Hence, high classification accuracy may partly reflect decoding of recording time rather than motor-imagery activity. The confusion matrix is consistent with this concern since errors are concentrated between alphabetically adjacent letters, such as \textbf{f/g}, rather than only between letters with clearly similar handwriting trajectories. If letters were recorded sequentially, adjacent letters would also be close in time, suggesting that the decoder may be learning temporal block structure instead of letter-specific motor-imagery patterns.

\begin{figure}[h!]
\centering
\includegraphics[width=0.9\textwidth]{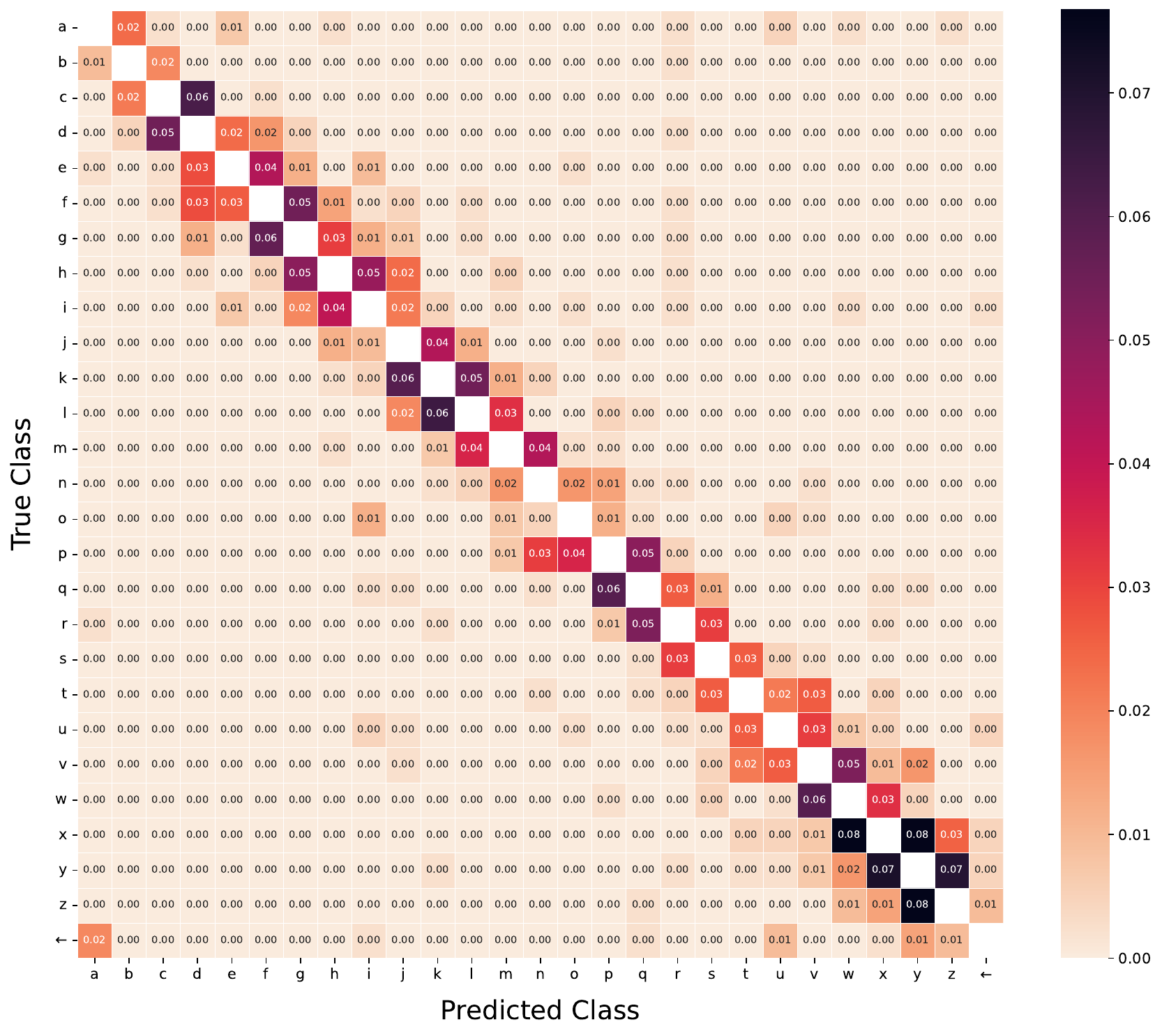}
\caption{Confusion matrix showing evidence consistent with temporal confounding in a block-designed EEG handwriting dataset. The main diagonal is omitted to make misclassifications more apparent.}
\label{fig:custom_montage}
\end{figure}

\section{\dset supplementary information}
\label{sec:lovw_supplementary}

\subsection{Synchronization}
\label{subsec:synchronization}

Three streams of data needed to be synchronized: The EEG stream from the amplifier, task cues from the recording computer, and the pen trajectories from the tablet. The LabStreamingLayer (LSL) networking ecosystem was used to send and receive the data between the devices. However, network latency can lead to streams being offset.

To ensure precise synchronization, we connected two photo-diodes to the amplifier, attaching one to a section of the tablet screen, and one to a section of the experimental monitor.  
The task program and the trajectory recording program were designed to flash the photo-diode region whenever a task cue was being shown, or a trajectory was drawn. These photodiode spikes were perfectly synchronized to the EEG since they fed into the same amplifier. The data/timestamps that were recorded with some latency by the LSL protocol were then aligned to the EEG stream using the closest photodiode spike in it.
We found network latencies and offsets of up to $80$ ms that were resolved using the photodiode-based synchronization.

\subsection{Montage and dataset details}

\begin{figure}[h!]
\centering
\includegraphics[width=0.4\textwidth]{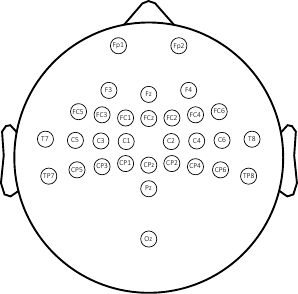}
\caption{A subset of the standard 10-10 montage designed to record more densely from the midline area, over the motor cortex.}
\label{fig:custom_montage}
\end{figure}

		

\begin{table}[h]
\centering
\caption{Details of data collection schedule. Sessions are ME unless otherwise noted.}
\label{tab:data_collection_sched}
  \vspace{2mm}
\renewcommand{\arraystretch}{1.15}
\setlength{\tabcolsep}{8pt}
\small



\begin{tabular}{cccc}
\hline
Session & Participant  & \# Trials & \# Days from first session \\
\hline
1 & P1      & 315  & 0   \\
2 & P1      & 696  & 34  \\
3 & P1      & 1268 & 346 \\
4 & P1 MI   & 1240 & 649 \\
5 & P1      & 120  & 659 \\
6 & P1 MI   & 400  & 659 \\
\hline
1 & P2      & 339  & 0   \\
2 & P2      & 336  & 2   \\
3 & P2      & 778  & 238 \\
4 & P2 MI   & 1198 & 308 \\
5 & P2      & 319  & 329 \\
\hline
1 & P3      & 474  &  0   \\
2 & P3 MI   & 300  &  0   \\
\hline
1 & P4      & 400  &  0  \\
2 & P4 MI   & 400  &  0   \\
\hline
\end{tabular}
\end{table}

\begin{table}[h]
\caption{Trials per setting across participants in \dset Dataset}
\label{tab:dataset_details}
\vspace{1mm}
\centering
\begin{sc}
\begin{tabular}{c c c c c}
\toprule
 & P1 & P2 & P3 & P4 \\
\midrule
\# ME Trials & 2398 & 1772 & 474 & 398 \\
\# MI Trials & 1640 & 1198 & 299 & 399 \\
\bottomrule
\end{tabular}
\end{sc}
\end{table}

\subsection{Permutation test}
\label{subsec:permutation_test}

Below we describe the exact non-parametric permutation test performed to conclude that subject-average performance of EEGNet was higher than REVE and CBraMod on \dset.

\textit{Null Hypothesis $H_0$}: EEGNet and FM architecture are equally likely to be correct on any given trial.

\textit{Test Statistic $T$}: Let

\[
f(\text{model}) := \frac{1}{n_s} \sum_s \bar{a}_s
\]

where $n_s$ is the number of subjects, and $\bar{a}_s$ is the mean (over trials) accuracy for subject $s$. Then

\[
T := f(\text{EEGNet}) - f(\text{FM})
\]

Note that the test statistic is not biased towards subjects with more trials than others. Given paired predictions per trial:

\begin{table}[h]
\centering
\begin{tabular}{ccc}
\hline
\textbf{Trial} & \textbf{EEGNet} & \textbf{FM} \\
\hline
1 & $\hat{y}_1^A$ & $\hat{y}_1^B$ \\
2 & $\hat{y}_2^A$ & $\hat{y}_2^B$ \\
$\vdots$ & $\vdots$ & $\vdots$ \\
$n$ & $\hat{y}_n^A$ & $\hat{y}_n^B$ \\
\hline
\end{tabular}
\end{table}

For each of 100,000 permutations, the predictions of EEGNet and FM are exchanged or maintained independently per trial with equal probability (i.e., a Bernoulli flip). The test statistic $T_i$ is computed on each permuted assignment, yielding a null distribution of $\{T_i\}_{i=1}^{100{,}000}$. The p-value is then estimated as

\[
p = \frac{\sum_i \mathbf{1}[T_i \geq T_{\text{obs}}]}{100{,}000}
\]

In the ME setting, after Bonferroni correction for multiple (2) comparisons, the null hypothesis was rejected for both CBraMod and REVE ($p < 0.00002$).
Thus, on the \dset dataset, the subject-average performance differences between architectures are statistically significant.


\subsection{Leave-One-Session-Out (LOSO) MI results}

In addition to LOSO ME, we also tested cross-session transfer in the MI setting, shown in Table \ref{tab:loso_MI_results}. However, decoding performance was not statistically significantly above chance.  

\begin{table}[h!]
  \caption{Leave-one-session-out accuracies for MI fixation-locked setting for P1 and P2.
           Values are reported as accuracy $\pm$ standard error (SE). No session showed performance significantly above chance.}
  \label{tab:loso_MI_results}
  \vspace{2mm}
  \centering
  \begin{small}
    \begin{sc}
      \begin{tabular}{lccccc}
        \toprule
        & \multicolumn{3}{c}{P1} & \multicolumn{2}{c}{P2} \\
        \cmidrule(lr){2-4} \cmidrule(lr){5-6}
        & Session 1 & Session 2 & Session 3 & Session 1 & Session 2 \\
        \midrule
        \# Trials & 800   & 440   & 400   & 798   & 400   \\
        Accuracy  & $0.278\pm0.016$ & $0.259\pm0.021$ & $0.263\pm0.022$
                  & $0.251\pm0.015$ & $0.280\pm0.022$ \\
        \bottomrule
      \end{tabular}
    \end{sc}
  \end{small}
\end{table}

\section{Supplementary experiments with prior dataset}

\subsection{Dataset description}
\label{sec:crell_dataset_description}

In this dataset from \cite{crell2024handwritten}, 20 healthy participants each performed cued air-writing movements corresponding to ten lowercase letters (a, d, e, f, j, n, o, s, t, v) using their index finger. EEG was recorded using a high-density montage while finger kinematics were simultaneously captured using an optical motion tracking system. The authors found that a subset of the letters (d,f,j,s,v) were more discriminable, so we restricted evaluation to this subset. 


\subsection{Prior dataset shows similar performance drop when movement onset is unknown}

\begin{figure}[h]
    \centering
    \includegraphics[width=0.4\textwidth]{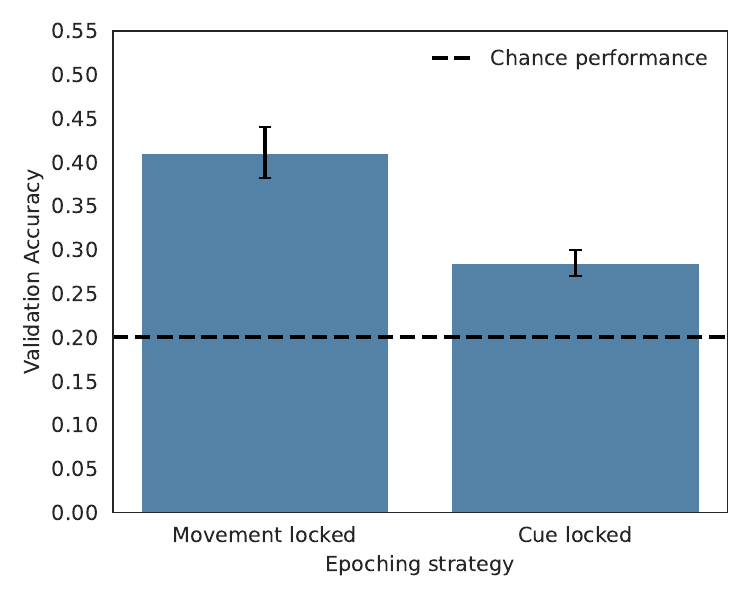}
    \caption{Effect of epoching strategy on decoding performance for the 5-letter classification. The chart shows the average accuracy and SE across the 20 subjects. Movement-locked setting achieves substantially higher validation accuracy than cue-locked epochs.}
    \label{fig:crell_performance_drop}
\end{figure}

Fig \ref{fig:crell_performance_drop} compares performance between movement locked epochs and cue locked epochs i.e where the onset is fixed relative to the start of the trial vs unknown. The cue-locked setting consistently yields lower performance across all subjects (shown in Table \ref{tab:crell_subjectwise_drop}), and is statistically significant (p < 0.001) using a Wilcoxon Signed-Rank test (N=20).

\begin{table}[h!]
\centering
\caption{Subject-wise classification scores for movement-locked and cue-locked epochs using EEGNet. Cue-locked setting exhibits uniformly worse performance across all subject.}
\label{tab:crell_subjectwise_drop}
  \vspace{2mm}
\renewcommand{\arraystretch}{1.15}
\setlength{\tabcolsep}{8pt}
\small

\begin{tabular}{ccc}
\hline
\textbf{Subject} & \textbf{Movement locked (accuracy $\pm$  SE) }& \textbf{Cue locked (accuracy $\pm$  SE)}\\
\hline
S01 & $0.255 \pm 0.0252$ & $0.244 \pm 0.0263$ \\
S02 & $0.413 \pm 0.0284$ & $0.270 \pm 0.0266$ \\
S03 & $0.470 \pm 0.0289$ & $0.261 \pm 0.0266$ \\
S04 & $0.505 \pm 0.0289$ & $0.478 \pm 0.0289$ \\
S05 & $0.423 \pm 0.0285$ & $0.301 \pm 0.0267$ \\
S06 & $0.410 \pm 0.0284$ & $0.174 \pm 0.0220$ \\
S07 & $0.361 \pm 0.0278$ & $0.281 \pm 0.0260$ \\
S08 & $0.367 \pm 0.0278$ & $0.254 \pm 0.0257$ \\
S09 & $0.361 \pm 0.0278$ & $0.231 \pm 0.0246$ \\
S10 & $0.448 \pm 0.0288$ & $0.331 \pm 0.0272$ \\
S11 & $0.760 \pm 0.0247$ & $0.344 \pm 0.0275$ \\
S12 & $0.315 \pm 0.0271$ & $0.285 \pm 0.0262$ \\
S13 & $0.333 \pm 0.0275$ & $0.303 \pm 0.0268$ \\
S14 & $0.346 \pm 0.0275$ & $0.270 \pm 0.0259$ \\
S15 & $0.268 \pm 0.0257$ & $0.220 \pm 0.0241$ \\
S16 & $0.700 \pm 0.0265$ & $0.386 \pm 0.0282$ \\
S17 & $0.380 \pm 0.0280$ & $0.215 \pm 0.0238$ \\
S18 & $0.410 \pm 0.0283$ & $0.309 \pm 0.0268$ \\
S19 & $0.447 \pm 0.0287$ & $0.296 \pm 0.0265$ \\
S20 & $0.247 \pm 0.0250$ & $0.246 \pm 0.0252$ \\

\hline
\textbf{Subject Avg.} & \textbf{0.411 $\pm$ 0.029} & $0.285 \pm 0.015$ \\
\hline
\end{tabular}
\end{table}

\subsection{Variability in Reaction Time (RT)}
\label{subsec:reaction_time_variability}

\begin{figure}[h]
    \centering
    \includegraphics[width=0.4\textwidth]{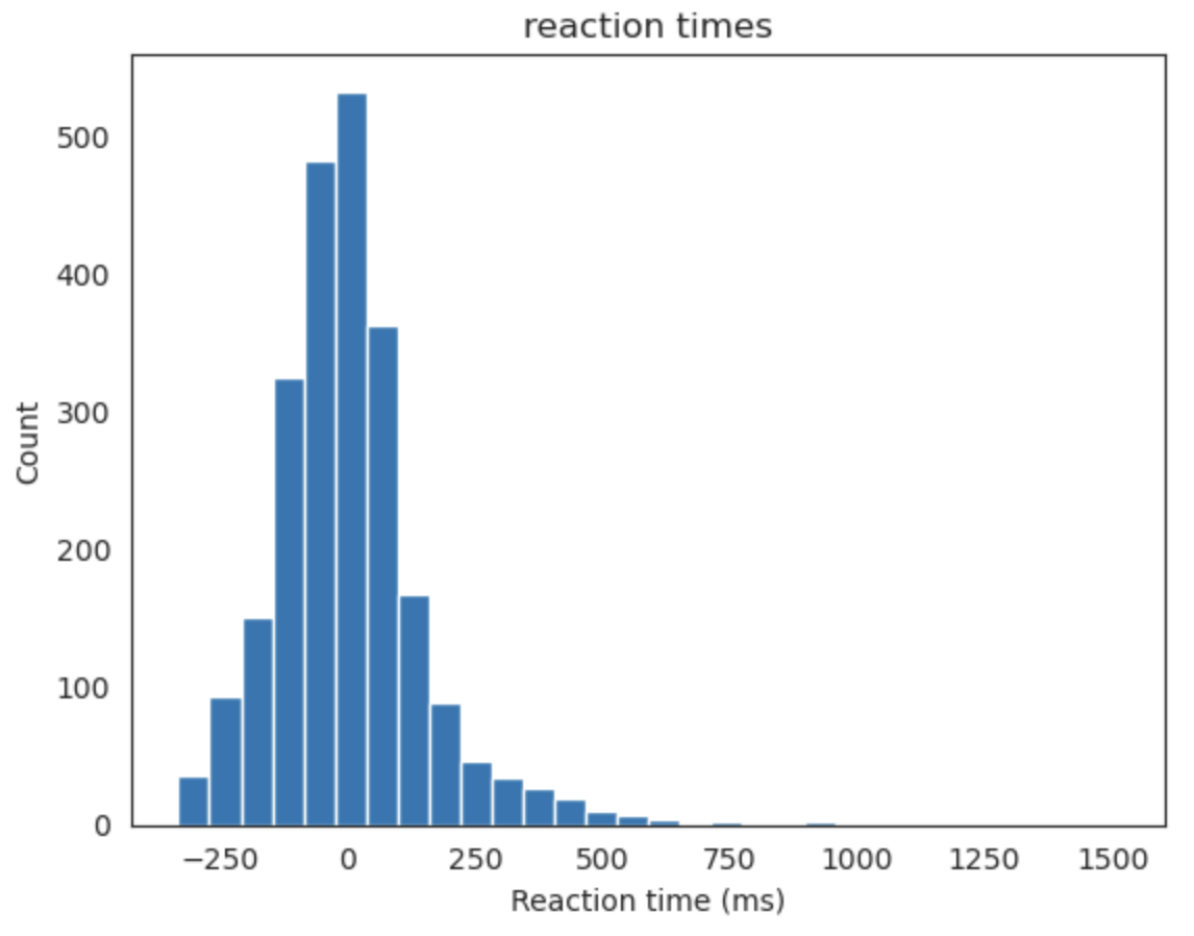}
    \caption{Mean-subtracted reaction time distribution indicates that cue-locked setting introduces jitter upto 250 ms compared to movement-locked setting.}
    \label{fig:reaction_times}
\end{figure}

To understand the differences in movement-locked vs cue-locked settings more, we analyzed the reaction times of a session from P1. The reaction time on any given trial is the time between the writing cue and the actual movement. The histogram of mean-subtracted reaction times for a subject indicates substantial variability upto $\pm$ 250 ms. This variability explains why epoching around the cue markers drastically reduces decoding accuracy, as it effectively introduces a per-trial temporal jitter of up to 250 ms relative to the true movement onset. Fig. \ref{fig:eegnet_jitter_plot} shows that performance drastically decreases as we inject artificial jitter into the movement-locked setting.

\begin{table}[h!]
\centering
\caption{Subject-wise decoding performance for REVE linear probing (LP), fine-tuning (FT), and EEGNet, reported as mean $\pm$ SE.}
\label{tab:crell_reve_vs_eegnet}
  \vspace{2mm}
\renewcommand{\arraystretch}{1.15}
\setlength{\tabcolsep}{8pt}
\small

\centering
\begin{tabular}{cccc}
\hline
\textbf{Subject} & \textbf{REVE LP} & \textbf{REVE FT} & \textbf{EEGNet} \\
\hline
\hline
S01 & $0.230 \pm 0.025$ & $0.229 \pm 0.027$ & $0.255 \pm 0.025$ \\
S02 & $0.357 \pm 0.011$ & $0.413 \pm 0.023$ & $0.413 \pm 0.028$ \\
S03 & $0.390 \pm 0.024$ & $0.400 \pm 0.026$ & $0.470 \pm 0.029$ \\
S04 & $0.369 \pm 0.099$ & $0.386 \pm 0.101$ & $0.505 \pm 0.029$ \\
S05 & $0.357 \pm 0.027$ & $0.370 \pm 0.017$ & $0.423 \pm 0.029$ \\
S06 & $0.317 \pm 0.022$ & $0.347 \pm 0.024$ & $0.410 \pm 0.028$ \\
S07 & $0.296 \pm 0.029$ & $0.312 \pm 0.030$ & $0.361 \pm 0.028$ \\
S08 & $0.420 \pm 0.041$ & $0.463 \pm 0.045$ & $0.367 \pm 0.028$ \\
S09 & $0.227 \pm 0.018$ & $0.261 \pm 0.017$ & $0.361 \pm 0.028$ \\
S10 & $0.482 \pm 0.032$ & $0.533 \pm 0.041$ & $0.448 \pm 0.029$ \\
S11 & $0.620 \pm 0.034$ & $0.657 \pm 0.037$ & $0.760 \pm 0.025$ \\
S12 & $0.263 \pm 0.026$ & $0.287 \pm 0.033$ & $0.315 \pm 0.027$ \\
S13 & $0.328 \pm 0.051$ & $0.346 \pm 0.046$ & $0.333 \pm 0.028$ \\
S14 & $0.354 \pm 0.026$ & $0.368 \pm 0.025$ & $0.346 \pm 0.028$ \\
S15 & $0.291 \pm 0.030$ & $0.294 \pm 0.025$ & $0.268 \pm 0.026$ \\
S16 & $0.557 \pm 0.027$ & $0.590 \pm 0.031$ & $0.700 \pm 0.027$ \\
S17 & $0.295 \pm 0.018$ & $0.344 \pm 0.024$ & $0.380 \pm 0.028$ \\
S18 & $0.388 \pm 0.043$ & $0.438 \pm 0.034$ & $0.410 \pm 0.028$ \\
S19 & $0.407 \pm 0.045$ & $0.427 \pm 0.040$ & $0.447 \pm 0.029$ \\
S20 & $0.288 \pm 0.024$ & $0.332 \pm 0.024$ & $0.247 \pm 0.025$ \\
\hline
\textbf{Subject Avg.} & $0.362 \pm 0.023$ & $0.390 \pm 0.024$ & \textbf{0.411 $\pm$ 0.029} \\
\hline
\end{tabular}
\end{table}

\section{Implementation supplementary information}

All experiments were conducted on a single NVIDIA A6000 GPU. SPaRCNet and EEGConformer were implemented using the Braindecode \cite{HBM:HBM23730} package. For the specialist models, a typical EEGNet training run took approximately 30 seconds per subject. Fine-tuning the EEG foundation models required approximately 1 hour across the 20-subject dataset, while preprocessing the full dataset required approximately 2 hours. These runtimes provide an approximate compute budget for reproducing the main experiments.

\end{document}